\documentclass[pre,twocolumn,english,showpacs,amsmath,amssymb,superscriptaddress,longbibliography]{revtex4-2}
\usepackage{color}
\usepackage{mathptmx}

\usepackage[T1]{fontenc}
\usepackage{float}
\usepackage{graphicx}
\usepackage[normalem]{ulem}
\usepackage{amsmath}
\usepackage{amssymb}
\usepackage{soul}
\usepackage{bm}
\usepackage{boldline,multirow}
\usepackage{comment}

\usepackage{bibunits}



\usepackage{babel}
\usepackage{bm}
\usepackage{xr}

\begin{document}



\renewcommand{\figurename}{\textbf{Fig.}}
\renewcommand{\thefigure}{\arabic{figure}}
\renewcommand{\thetable}{\arabic{table}}

\title{Microscopic understanding of ion solvation in water}

\author{Rui Shi}
\email{ruishi@zju.edu.cn}
\affiliation{Zhejiang Province Key Laboratory of Quantum Technology and Device, Department of Physics, \\
Zhejiang University, Zheda Road 38, Hangzhou 310027, China.}
\affiliation{Department of Fundamental Engineering, Institute of Industrial Science, \\
University of Tokyo, 4-6-1 Komaba, Meguro-ku, Tokyo 153-8505, Japan}
\author{Anthony J. Cooper}
\affiliation{Department of Fundamental Engineering, Institute of Industrial Science, \\
	University of Tokyo, 4-6-1 Komaba, Meguro-ku, Tokyo 153-8505, Japan}
\affiliation{Present address: Department of Physics, University of California, Santa Barbara, CA 93106-9530, U.S.A.} 
\author{Hajime Tanaka}
\email{tanaka@iis.u-tokyo.ac.jp}
\affiliation{Department of Fundamental Engineering, Institute of Industrial Science, \\
University of Tokyo, 4-6-1 Komaba, Meguro-ku, Tokyo 153-8505, Japan}
\affiliation{Research Center for Advanced Science and Technology, University of Tokyo, 4-6-1 Komaba, Meguro-ku, Tokyo 153-8505, Japan}

\date{\today}
\vspace{-7mm}
\begin{abstract}
Solvation of ions is ubiquitous on our planet. Solvated ions have a profound effect on the behavior of ionic solutions, which is crucial in nature and technology. Experimentally, ions have been classified into ``structure makers'' or ``structure breakers'', depending on whether they slow down or accelerate the solution dynamics. Theoretically, the dynamics of ions has been explained by a dielectric friction model combining hydrodynamics and charge-dipole interaction in the continuum description. However, both approaches lack a microscopic structural basis, leaving the microscopic understanding of salt effects unclear. Here we elucidate unique microscopic features of solvation of spherical ions by computer simulations. We find that increasing the ion electric field causes a sharp transitional decrease in the hydration-shell thickness, signaling the ion mobility change from the Stokes to dielectric friction regime. The dielectric friction regime can be further divided into two due to the competition between the water-water hydrogen bonding and ion-water electrostatic interactions: Whether the former or latter prevails determines whether the water dynamics are accelerated or decelerated. In the ion-water interaction predominant regime, a specific combination of ion size and charge stabilizes the hydration shell via orientational-symmetry breaking, reminiscent of the Thomson problem for the electron configuration of atoms. Notably, the hydration-shell stability is much higher for a composite coordination number than a prime one, a prime-number effect on solvent dynamics. These findings are fundamental to the structure breaker/maker concept and provide new insights into the solvent structure and dynamics beyond the continuum model, paving the way towards a microscopic theory of ionic solutions. 
\end{abstract}

\maketitle

The ability of water to dissolve salts provides the basic environment for many chemical, biological, geological and technological processes, which is crucial for our life. The presence of ions specifically affects the structure and kinetics of water~\cite{gurney1953ionic,nightingale1959phenomenological,ball2008water,bakker2008structural,marcus2009effect}, which further impact a broad class of phenomena, such as cloud formation~\cite{hudait2014ice}, protein function~\cite{bellissent2016water,mukherjee2019mechanism}, ice nucleation~\cite{koop2000water,soria2018simulation,conde2018molecular}, gas capture~\cite{liu2013simulations}, interfacial organization~\cite{gonella2021water}, and charge transport~\cite{peng2018effect} in our planet. Despite intensive studies and the accumulation of experimental data of salt solutions over centuries, the microscopic mechanism behind the specific ionic effects has remained poorly understood so far.

The solvation of ions alters the structure and dynamics of water. Empirically, it is well known that the viscosity $\eta$ of a salt solution can be described by the Jones-Dole equation~\cite{jones1929viscosity}:
\begin{equation}
\eta(c)/\eta_0=1+Ac^{1/2}+Bc,
\label{eq:jd}
\end{equation}
where $\eta_0$ is the viscosity of pure water, and $c$ is the salt concentration. This equation applies to a broad class of ions and has been linked to the Hofmeister series classification of ions as the structure-making type for $B>0$ or the structure-breaking type for $B<0$~\cite{gurney1953ionic,marcus2009effect}. Thermodynamic measurements support this classification: the $B$-coefficient correlates with the ionic entropy, a measure of the degree of ion-induced order in an aqueous solution~\cite{gurney1953ionic}. Despite the lack of direct structural characterization, these observations suggest a fundamental connection between the structural and dynamic effects of ions: the structure-making ion promotes water hydrogen-bond (H-bond) structure, thus slowing down the dynamics, whereas the structure-breaking one destroys the water structure, thus accelerating water dynamics~\cite{ball2008water}. Such an idea has become one of the most common languages for understanding ion effects in salt solutions. However, the situation is not that simple. Recent neutron scattering experiments, which measure the two-body density correlation functions, have detected ion-induced distortions of water structure for both K$^+$ and Na$^+$ ions and thus identified these ions as structure breakers~\cite{mancinelli2007hydration}. On the other hand, viscosity measurements have classified K$^+$ and Na$^+$ as structure breaker and maker, respectively, based on the sign of the $B$-coefficient~\cite{marcus2009effect}. Such a discrepancy has been supported computationally in a salt model~\cite{gallo2011ion}, yet whose origin has remained an open question. This may be because the microscopic characterization of the water structure around ions has been limited to the two-body-level information. 
We stress that many-body correlations play an essential role in local structuring ordering in liquids, including water~\cite{tanaka2019revealing}. 

Theoretically, ion solvation has been discussed mainly from the viewpoint of the ion dynamics in a solution. A spherical particle in a fluid experiences hydrodynamic friction oppositive the direction of motion. The hydrodynamic friction obeys Stokes' law, derived from the Navier-Stokes equation at the low Reynolds-number limit. When a charged particle is immersed in a dielectric medium consisting of dipolar molecules, the charge (i.e., monopole) electrostatically interacts with dipoles, and dipoles react to the motion of a charge. The delayed reaction of water dipoles to ion motion effectively exerts dielectric friction in addition to the hydrodynamic friction. This seminal idea pioneered by Born~\cite{born1920} has been successfully applied to salt solutions and developed by Boyd~\cite{boyd1961}, Zwanzig~\cite{zwanzig1970dielectric}, Onsager~\cite{hubbard1977dielectric}, Wolynes~\cite{wolynes1978} and their coworkers. In the dielectric friction model, the total friction $\zeta$ experienced by a spherical ion with charge $q$ and radius $R$ is given by
\begin{equation}
	\zeta = f_1 \eta R + f_2 \left(\frac{\epsilon_0-\epsilon_\infty}{\epsilon_0^2} \right) \frac{\tau_D q^2 e^2}{R^3}.
	\label{eq:df}
\end{equation}

The first and second terms on the right side of this equation correspond to the hydrodynamic and dielectric friction, respectively. $f_1$ and $f_2$ are constants related to the boundary condition (stick or slip), $\eta$ is the viscosity of the medium, $\epsilon_0$ and $\epsilon_\infty$ are the low-frequency- and high-frequency-limit dielectric constants, respectively, $\tau_D$ is the dielectric relaxation time of the medium, and $e$ is the elementary charge. The dielectric friction theory predicts that large ions follow Stokes' law ($\zeta \propto R$), whereas for small ions, the dielectric friction ($\zeta \propto R^{-3}$) is dominant. For example, this theory can explain the slower diffusion of Li$^+$ than Na$^+$ in water even though Li$^+$ is smaller than Na$^+$. Although the dielectric friction concept successfully incorporates ion-dipole interactions with hydrodynamics and captures the essential physics of solvation, it treats the solvent as a continuum medium, neglecting the microscopic aspects of solvent-solvent (water-water H-bonding) and ion-solvent (monopole-dipole) interactions. Therefore, the theory suffers from intrinsic difficulties in describing the solvent dynamics and the specificity of salt effects in aqueous solutions. We note that a continuum description is quite helpful to understand macroscopic phase behaviors such as phase separation~\cite{onuki2011solvation}.

This article studies the solvation of a spherical ion in liquid water and its effect on solvent dynamics at the microscopic level by molecular dynamics simulations (see supplementary material for the details). We utilized a newly developed non-polarizable force field~\cite{zeron2019} for a series of aqueous solutions based on a realistic TIP4P/2005 water model~\cite{abascal2005} that has been shown to accurately describe the structure and dynamics of liquid water~\cite{vega2009}. Starting from this potential, we continuously modified the van der Waals (VDW) radius $R$ and charge $q$ of cation while keeping the force field parameters fixed for the anion and TIP4P/2005 water. We have performed a high-throughput computational scanning of 1332 model cations with different sizes and charges, covering the alkali metal ions, the alkaline metal ions, and the IIIA-group metal ions. This allows us to systematically study the salt effects in aqueous solutions with microscopic details beyond the continuum theory. We emphasise here that continuous change of $R$ and $q$ is crucial for revealing a transitional nature of the solvation state. Although the dispersion energy may affect the salt effects as well~\cite{andreev2017coarse,andreev2018influence}, here we focus on the effects of ion size and charge that directly determine the intensity of ion electric field and thus the strength of dielectric friction. To minimize ion-ion interactions such as ion pairing~\cite{van2016water} and cooperativity effects~\cite{tielrooij2010cooperativity} that were activated in cencentrated solutions, here we constrain our study in a dilute concentration ($\sim0.16$ mol/kg), focusing on ion-water interactions.

From the high-throughput scanning of $R$ and $q$, we have successfully revealed two sequential structural transformations of the water hydration shell with an increase in the ion electric field. An increase in the monopole-dipole interaction induces the first transformation from a thick to a thin hydration shell. Its further increase leads to the second transformation from radially disordered to ordered alignments of water dipoles in the hydration shell, caused by the monopole-dipole interaction overwhelming the water-water H-bonding and dipole-dipole interaction inside the hydration shell. Furthermore, in the regime of radial dipolar alignment, the dipolar-dipolar repulsion between water molecules in a spherical shell leads to the high (low) orientational symmetry for a shell with a composite (prime) number of water molecules inside, stabilizing (destabilizing) the water hydration shell. We have revealed that these hydration shell structures control ion dynamics and solvation stability. We have also shown that our model nicely explains the experimentally observed dynamic behaviors of aqueous ionic solutions.

\begin{figure*}[t!]
	\begin{center}
		\includegraphics[width=14cm]{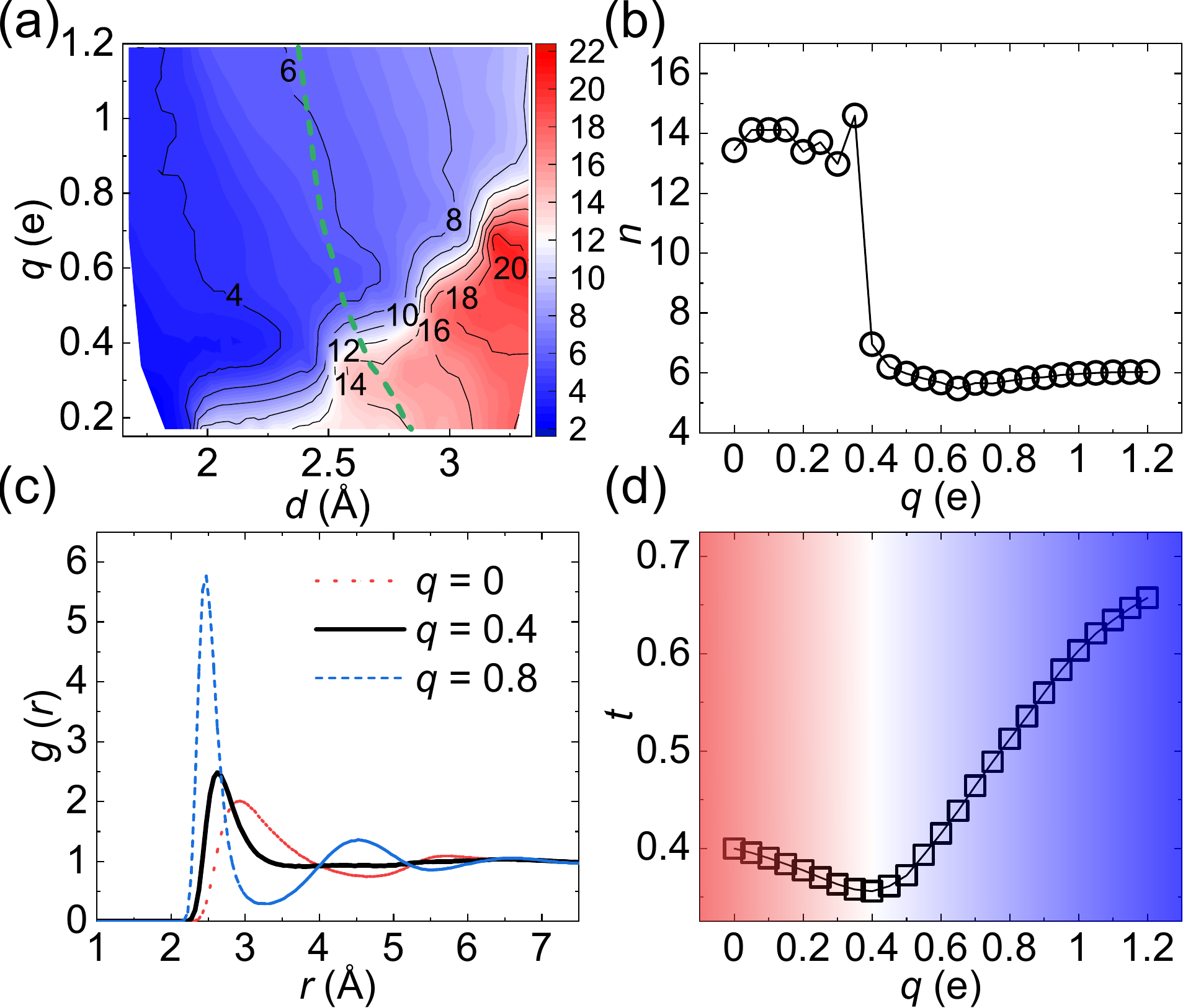}
	\end{center}
	\vspace{-7mm}
	\caption{Structural transition of the hydration shell. We measure the the number of water molecules in the hydration shell, $n$, and the 
		radius of the hydration shell, $d$, around cations of charge $q$. The radius $d$ is defined as the first peak position in the cation-oxygen radial distribution function (RDF). 
		(a) The contour map of $n$ shown as a function of $d$ and $q$.  The color bar represents the value of $n$. (b-d) The cationic coordination number $n$ (b), the cation-oxygen RDF $g(r)$ (c), and the translational order parameter $t$ of the cationic hydration shell (d) along the green dashed line in panel a, along which the VDW size of the cation is kept constant. In (b) and (d), the cationic charge is increased from $q=0$ to $q=1.2~e$ with a step of $0.05~e$.}
	\vspace{-3mm}
	\label{fig:transition}
\end{figure*}

\subsection*{Structural transition of hydration shell: translational ordering}
A cation attracts water molecules via the Coulomb interaction to form (layered) spherical shells of solvent molecules around it. The spherical shells are usually called hydration shells, which are observed as peaks in the cation-oxygen radial distribution function (RDF), $g(r)$. Hereafter, we specifically use the term ``hydration shell'' to refer to the first hydration shell closest to the cation, which is most strongly influenced by the solvation of ion~\cite{omta2003negligible}. The radius $d$ of the hydration shell is determined as the first peak position of $g(r)$, and the coordination number $n$ is defined as the number of water molecules in the hydration shell of the cation. 
Hereafter we use the hydration shell radius $d$ instead of the VDW radius $R$ to characterize the ion size since $d$ is experimentally accessible and directly determines the interaction between the ion and water molecules in the hydration shell.
Figure~\ref{fig:transition}(a) shows the contour map of $n$ for cations with charge $q \leqslant 1.2~e$. We can see that a cation with a small charge typically has a large coordination number $n>12$. However, $n$ dramatically drops to 4-8 as $q$ increases, suggesting a sharp structural transformation of the hydration shell. Along an iso-ion-size line (the green dashed curve in Fig.~\ref{fig:transition}(a), on which the VDW radius of cation $R$ is kept constant), $n$ shows a sharp drop from $\sim14$ to $\sim6$ at a transition point $q_s=0.4~e$ with increasing $q$ (Fig.~\ref{fig:transition}(b)).  Figure~\ref{fig:transition}(c) and Figs.~\ref{fig:rdfq}-\ref{fig:rdfq1} show the typical cation-oxygen RDF, $g(r)$, along the iso-ion-size line. Below $q_s$, the cation has a broad yet asymmetric hydration shell peak, whereas, above it, a narrow, symmetric hydration-shell peak is formed around the cation. Moreover, the second peaks are out-of-phase between cations with $q>q_s$ and $q<q_s$, and the peaks other than the first peak fade away just around the transition, $q_s=0.4~e$. Such behavior is also captured by a translational order parameter $t$ of the solvent~\cite{errington2001} (see supplementary material). As shown in Fig.~\ref{fig:transition}(d), as $q$ increases, $t$ first decreases and then increases substantially, showing a minimum at the transition point $q_s$, in agreement with the structural features shown in Fig.~\ref{fig:transition}(c).

Not only the ion-water distance but also the water-ion-water angle in the hydration shell are regulated to increase the degree of order as the ionic charge increases (Fig.~\ref{fig:angleq}). To understand the mechanism of such behavior, we have simulated an ensemble of 50 non-equilibrium trajectories. Started from 50 independent configurations equilibrated for $q=0.17~e < q_s$, we first jump the value of $q$ instantaneously to $0.68~e > q_s$ at time 0 and then monitored the structural evolution with time. The results are shown in Figs.~\ref{fig:theta} and \ref{fig:rdfnhb}. We can see that the structural transformation starts with the orientational ordering of water dipole in the hydration shell along the radial direction at a time scale of $\sim 0.03$~ps. This process is accompanied by the breakdown of H-bonds and the bifurcation of the first $g(r)$ peak of the hydration shell. This time scale corresponds to the librational mode of a water molecule. The rotational time correlation function in bulk water indicates that the libration mode couples with a reorientation of $\sim 23~^\circ$ of water dipole at $\sim 0.03$~ps (Fig.~\ref{fig:theta}), leading to the breakdown of H-bonds in water. Thus, we reveal that the coupling between ion-induced dipolar ordering and the libration mode breaks the H-bonds and induces the structural reorganization of the hydration shell.

This breakdown of H-bonds by dipolar ordering is critical for another structural transition in the hydration shell (see below).
These results demonstrate that the solvation of a (nearly) neutral particle is intrinsically different from that of an ion.
Below the transition ($q<q_s$), an ion is solvated by a large number of solvent molecules with disordered arrangements. Since the ion does not perturb the solvent structure strongly, its dynamics can be described by Stokes law.  Above the transition, the dielectric friction mechanism originating from dipolar ordering should play a critical role in ion dynamics. Therefore, we consider that the regimes below and above the transition should correspond to the Stokes and dielectric friction regimes, respectively~\cite{martelli2012varying}.

\subsection*{Dynamic crossover}

\begin{figure*}[t!]
	\begin{center}
		\includegraphics[width=16cm]{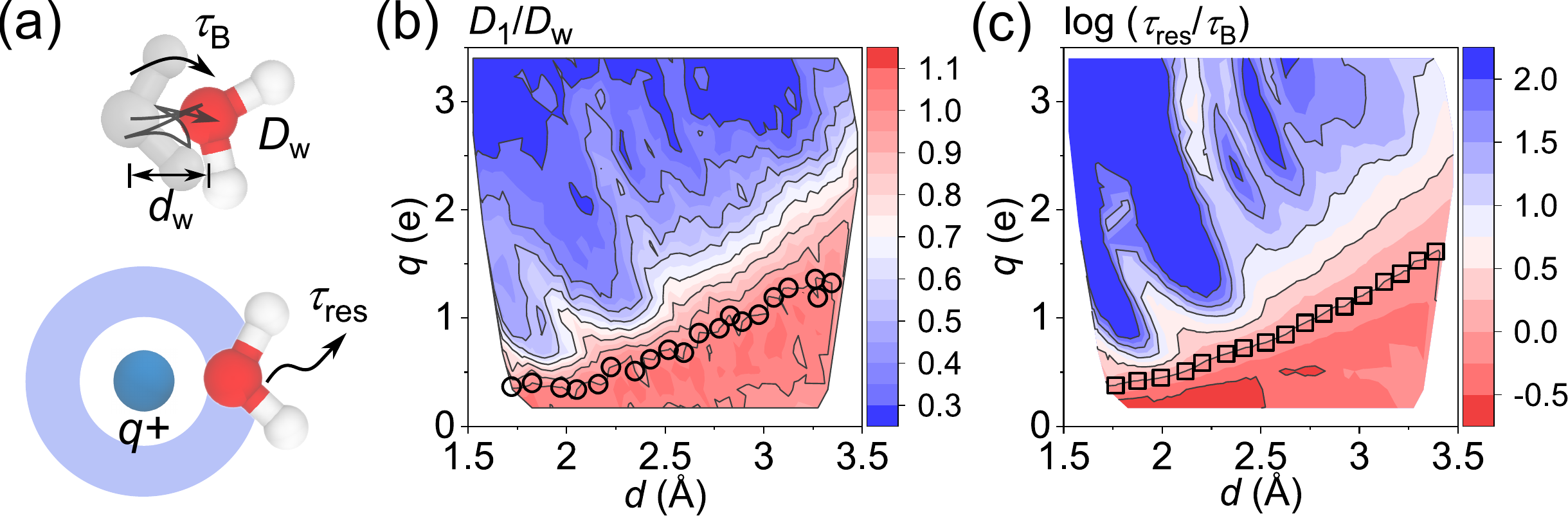}
	\end{center}
	\vspace{-7mm}
	\caption{Crossover of solvent dynamics near a spherical ion. (a) Schematic illustration of the definition of $\tau_\mathrm{B}$ and $\tau_\mathrm{res}$. The Brownian time $\tau_\mathrm{B}$ is defined as the time for a water molecule to diffuse over a distance of its diameter: $\tau_\mathrm{B}=d_\mathrm{w}^2/6D_\mathrm{w}$. Here $D_\mathrm{w}$ is the diffusion coefficient of a water molecule in bulk, and $d_\mathrm{w}$ is the diameter of a water molecule that is obtained from the first peak position of the oxygen-oxygen RDF of bulk water. The residence time $\tau_\mathrm{res}$ is defined as the characteristic time for the hydrated water to leave the hydration shell of a cation (see supplementary material). (b) The ratio $D_1 / D_\mathrm{w}$ in the $q$-$d$ plane. Here, $D_1$ is the diffusion coefficient of a water molecule in the vicinity of the hydration shell of a cation (see supplementary material). The color bar represents the value of $D_1 / D_\mathrm{w}$. (c) The logarithm of the ratio $\tau_\mathrm{res} / \tau_\mathrm{B}$, $\log (\tau_\mathrm{res} / \tau_\mathrm{B})$, plotted in the $q$-$d$ plane. The color bar represents the value of $\log (\tau_\mathrm{res} / \tau_\mathrm{B})$. In (b) and (c), the circle and square symbols correspond to the dynamic crossover lines, for whose ($q$, $d$) $D_1 = D_\mathrm{w}$ and $\tau_\mathrm{res} = \tau_\mathrm{B}$ are satisfied, respectively.}
	\vspace{-3mm}
	\label{fig:dynamics}
\end{figure*}

\begin{figure*}[t!]
	\begin{center}
		\includegraphics[width=12cm]{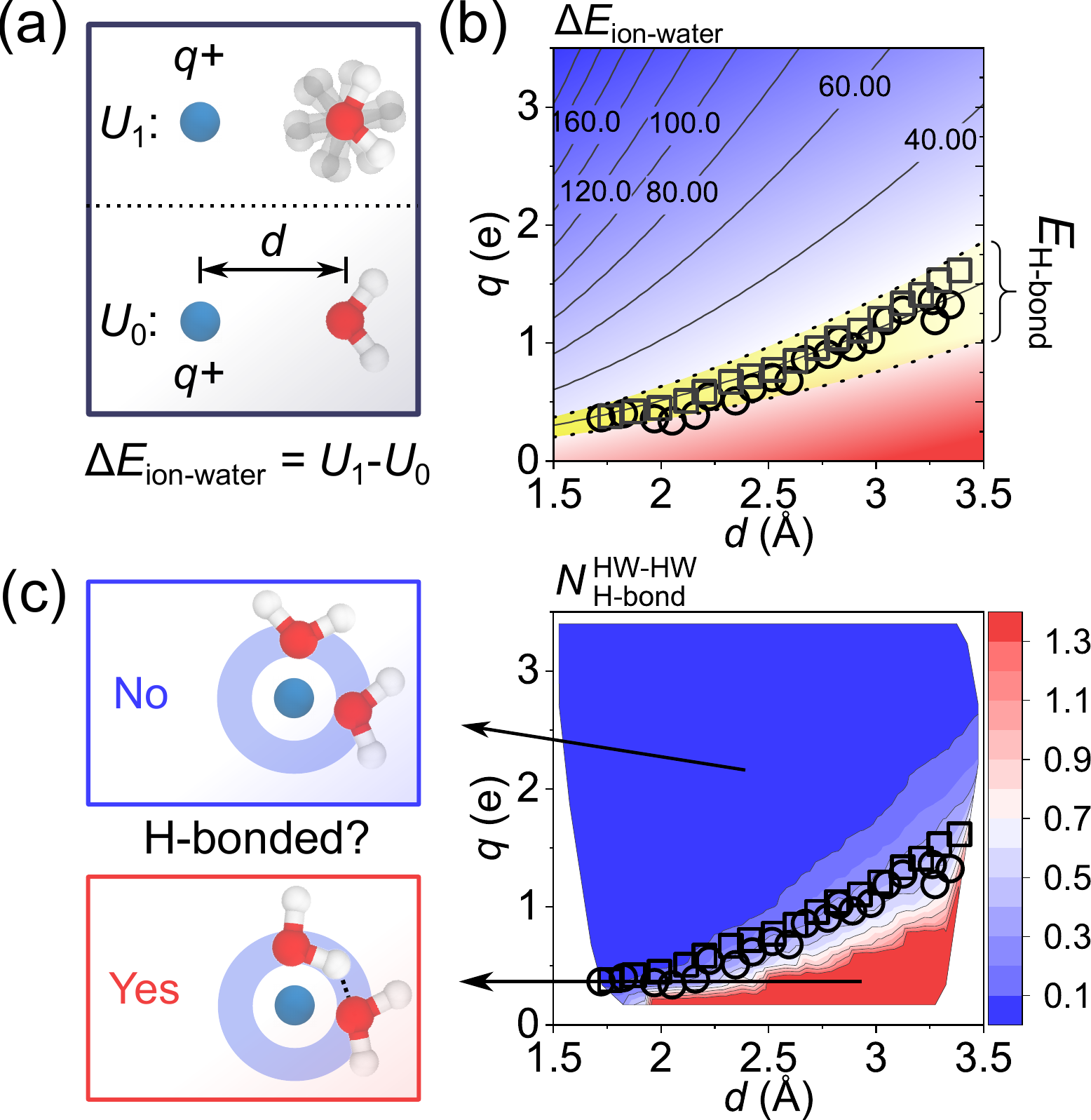}
	\end{center}
	\vspace{-7mm}
	\caption{Ion-water interaction versus water-water hydrogen bonding upon solvation. (a) Schematic illustration of the definition of the excess ion-water interaction energy $\Delta E_\mathrm{ion-water} \left(q,d\right) = U_1 - U_0$, where $U_1$ is the interaction energy between an ion and a randomly oriented water molecule, and $U_0$ is the interaction energy between an ion and a water molecule in the lowest-energy water orientation, for a given charge of an ion $q$ and a given ion-oxygen distance $d$. (b) The contour map of $\Delta E_\mathrm{ion-water}$ (in the unit of kJ/mol). The dynamic crossover lines (symbols, reproduced from Fig.~\ref{fig:dynamics}) coincide well with the water-water H-bonding energy $E_\mathrm{H-bond}$ (yellow band) in bulk water. The width of the yellow band denotes the range of $E_\mathrm{H-bond}$ under thermal fluctuations (Fig.~\ref{fig:hbenergy}). (c) The contour map of the average number of H-bonds $N_\mathrm{H-bond}^\mathrm{HW-HW}$ between a hydrated water molecule and other water molecules within the same hydration shell. The color bar represents the value of $N_{\rm H-bond}^{\rm HW-HW}$. The dynamic crossover lines (symbols) coincide well with the crossover from H-bonded to non-H-bonded hydration shell. The left two panels are schematic illustrations of H-bonded and non-H-bonded hydration shells of a cation.}
	\vspace{-3mm}
	\label{fig:hbond}
\end{figure*}

The solvent dynamics should be controlled by the size and charge of ions. Figures~\ref{fig:dynamics}(b) and (c) show the diffusion coefficient $D_1$ and the residence time $\tau_\mathrm{res}$ of hydrated water molecules (see Fig.~\ref{fig:dynamics}(a) and supplementary material for the explanation). Clearly, the dynamics of hydrated water molecules is slower under a stronger electrical field, i.e., for cations with larger $q$ and smaller $d$. 
Here we scale $D_1$ and $\tau_\mathrm{res}$ by the diffusion coefficient of bulk water $D_\mathrm{w}$ and the Brownian time of bulk water $\tau_\mathrm{B}$ (the time for a water molecule in bulk to diffuse over a distance of its diameter), respectively. Then, the two scaled variables exhibit similar dynamic crossover behaviors from the accelerated to decelerated dynamics in the $q-d$ diagram (compare Figs.~\ref{fig:dynamics}(b), (c)). This can be seen clearly in Fig.~\ref{fig:hbond}(b), which plots the dynamic crossover lines ($D_1 = D_\mathrm{w}$ and $\tau_\mathrm{res} = \tau_\mathrm{B}$) in the $q-d$ diagram, together with a contour map of the ion-water interaction energy $\Delta E_\mathrm{ion-water} (q,d)$ (see Fig.~\ref{fig:hbond}(a) for the explanation of $\Delta E_\mathrm{ion-water}$). Remarkably, the dynamic crossover lines of the two dynamic quantities almost coincide with each other in the $q-d$ diagram, suggesting a common physical mechanism behind their relative behaviors of the two quantities.

In bulk water, water molecules tend to form an H-bond network, whose strength $E_\mathrm{H-bond}$ determines the dynamics of water. Adding ions to water introduces ion-water interaction $\Delta E_\mathrm{ion-water} (q,d)$, which competes with the water-water H-bonding to form hydration shells around ions. The hydrated water under these competing interactions can have dynamics distinct from the bulk water. If $\Delta E_\mathrm{ion-water} > E_\mathrm{H-bond}$, the hydrated water should be dominated by the strong ion-water interaction rather than water-water H-bonding, leading to strong binding of water molecules to ions while breaking H-bonds (see Fig.~\ref{fig:hbond}(c)). This strong water binding to ions should slow down the ion motion, i.e., decelerated ion dynamics. On the other hand, if $\Delta E_\mathrm{ion-water} < E_\mathrm{H-bond}$, the hydration energy should be too weak to compete with the water-water H-bonding, but the presence of ions should still perturb the water's H-bond network. Although hydrated water molecules may remain H-bonded (Fig.~\ref{fig:hbond}(c)), the average H-bond number and the H-bond strength are both reduced statistically (Fig.~\ref{fig:nhb}). Thus, the hydrated water tends to escape from the ion and form new H-bonds with bulk water, leading to accelerated water dynamics. Based on this consideration, we compare the dynamic crossover line with the condition of $\Delta E_\mathrm{ion-water} = E_\mathrm{H-bond}$ and find that they coincide well with each other within errors (see Fig.~\ref{fig:hbond}(b)). Moreover, the dynamic crossover line also coincides with the structural transformation from an H-bonded to a non-H-bonded hydration shell (Fig.~\ref{fig:hbond}(c)). This finding encourages us to introduce a new length $\lambda_\mathrm{HB} (q)$, which satisfies $\Delta E_\mathrm{ion-water} \left[d=\lambda_\mathrm{HB}(q)\right] = E_\mathrm{H-bond}$. This characteristic length $\lambda_\mathrm{HB} (q)$ over which water behaves water-like relatively free from ions naturally explains the dynamic crossover in aqueous solutions with spherical ions. This length $\lambda_\mathrm{HB} (q)$ may play a crucial role in our understanding of solvated water, as the Debye length $\lambda_{\rm D}$ and the Bjerrum length $\lambda_{\rm B}$ do in the charge-related solution problem.

\subsection*{Orientational symmetry breaking of the hydration shell: Polyhedral ordering}
\begin{figure*}[t!]
	\begin{center}
		\includegraphics[width=16cm]{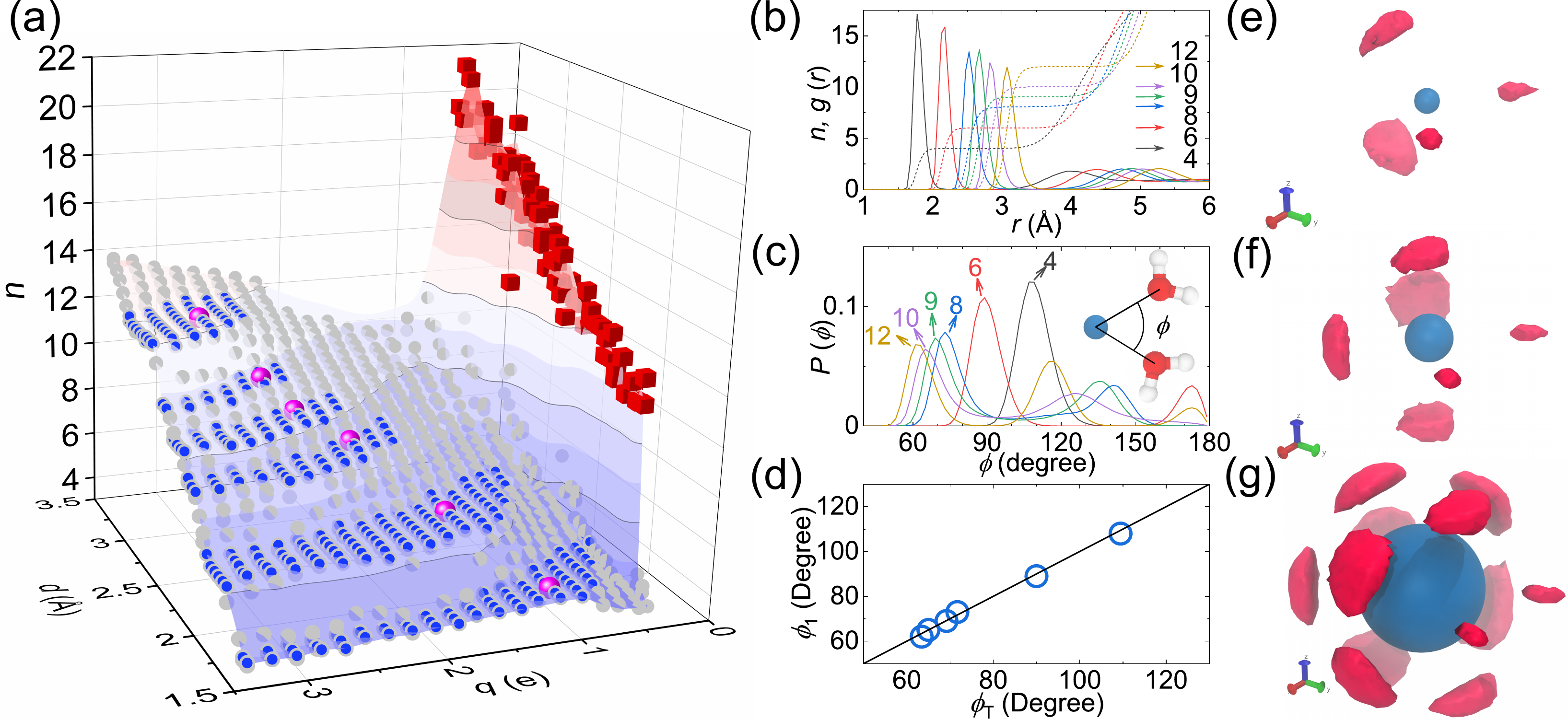}
	\end{center}
	\vspace{-7mm}
	\caption{Orientational symmetry breaking of the hydration shell. (a) The coordination number $n$ of a cation plotted in the $q$-$d$ plane. The plateaus of $n$ highlighted by blue spheres indicate the formation of stable hydration shells. (b-c) The cation-oxygen RDF $g(r)$ (b) and the distribution of the oxygen-cation-oxygen angle $P(\phi)$ (c) for typical cations in the plateau region of $n=4,6,8,9,10,~\mathrm{and}~12$ (big magenta spheres in (a)). The dash lines in (b) represent the coordination number of cations. The inset in (c) illustrates the definition of $\phi$ as the angle formed between the cation and two water oxygen atoms in the hydration shell. The arrows in (b) and (c) denote the corresponding coordination number $n$ of the cation. (d) The relationship between the smallest angles $\phi_1$ for the primary peak of $P(\phi)$ and the solution of the Thomson problem, $\phi_\mathrm{T}$~\cite{erber1991}. The line has a slope of 1, suggesting $\phi_1=\phi_\mathrm{T}$. (e-g) Spatial distribution of water molecules around a central ion with tetrahedral (e), octahedral (f), and icosahedral (g) symmetries. The size of the central ion does not represent its actual size. }
	\vspace{-3mm}
	\label{fig:symmetry}
\end{figure*}

For $d<\lambda_\mathrm{HB}(q)$, we can see the nonmonotonic dependence of the dynamic properties on $q$ and $d$ (Figs.~\ref{fig:dynamics}(b), (c)), e.g., 
specific regions with ultralong residence time ($\tau_\mathrm{res} / \tau_\mathrm{B} \geqslant 10^3$) in the $q-d$ diagram. 
This suggests the formation of a stable hydration shell (Fig.~\ref{fig:dynamics}(c)) for a specific combination of $q$ and $d$. 
Further analysis reveals that the regions of stable hydration correspond to the plateaus of the coordination number $n$ (Fig.~\ref{fig:symmetry}(a)). 
In the $q-d$ range of this study, we find that such plateaus emerge only when $n$ is a composite number, i.e., $n=4,6,8,9,10,~\mathrm{and}~12$, whereas the hydration structure of a prime coordination number is unstable, or even hard to form. A similar prime-number effect has been reported in the Thomson problem that considers the most stable configuration of $n$ point charges on a sphere with an opposite charge: In the range of $n \leqslant 30$, Glasser and Every found that configurations with prime $n$ show instability, whereas those with composite $n$ are stable~\cite{glasser1992}. The similarity of ion hydration to the Thomson problem can be explained by the fact that water dipoles in the hydration shell should align radially under strong enough electrostatic interaction from the central ion and repel each other on the ion surface, similar to the behavior of point charges of the same sign constrained on a spherical shell.

The emergence of the stable hydration shell with composite $n$ is associated with the orientational symmetry breaking of the hydration shell. Figure~\ref{fig:symmetry}(b) plots the cation-oxygen RDFs for typical cations in the plateau region (see big magenta spheres in Fig.~\ref{fig:symmetry}(a)). Every RDF shows a sharp first peak and a well-separated second shell (the first minimum goes to zero), indicating the local breaking of translational symmetry in the stable hydration shell. Moreover, we find that the spatial distributions of hydrated water molecules exhibit unique orientational symmetries, such as tetrahedral ($n=4$), octahedral ($n=6$) and icosahedral ($n=12$) symmetry in the plateau regions (Figs.~\ref{fig:symmetry}(e), (f) ,(g)). These hydration shell structures are markedly different from the nearly homogeneous shell structures in the non-plateau regions (Fig.~\ref{fig:sdf}). We also show the distribution of the oxygen-cation-oxygen angle $\phi$ in Fig.~\ref{fig:symmetry}(c). As $n$ increases, the smallest angle $\phi_1$ (i.e., the primary peak position of $P(\phi)$) takes characteristic values that follow the solution $\phi_\mathrm{T}$ of the Thomson problem (Fig.~\ref{fig:symmetry}(d)). The local orientational symmetry breaking in the plateau region at composite number $n$ explains the high stability of the hydration shell, the ultralong residence time of water in the hydration shell, and the slow solvent dynamics in the plateau regions (Fig.~\ref{fig:dynamics}(b),(c)). We emphasise that the continuum theory can hardly capture this unique behavior originating from the molecular-level water-water interactions under the electric field of ions.

\subsection*{Comparison with experiments}
\begin{figure*}[t!]
	\begin{center}
		\includegraphics[width=16cm]{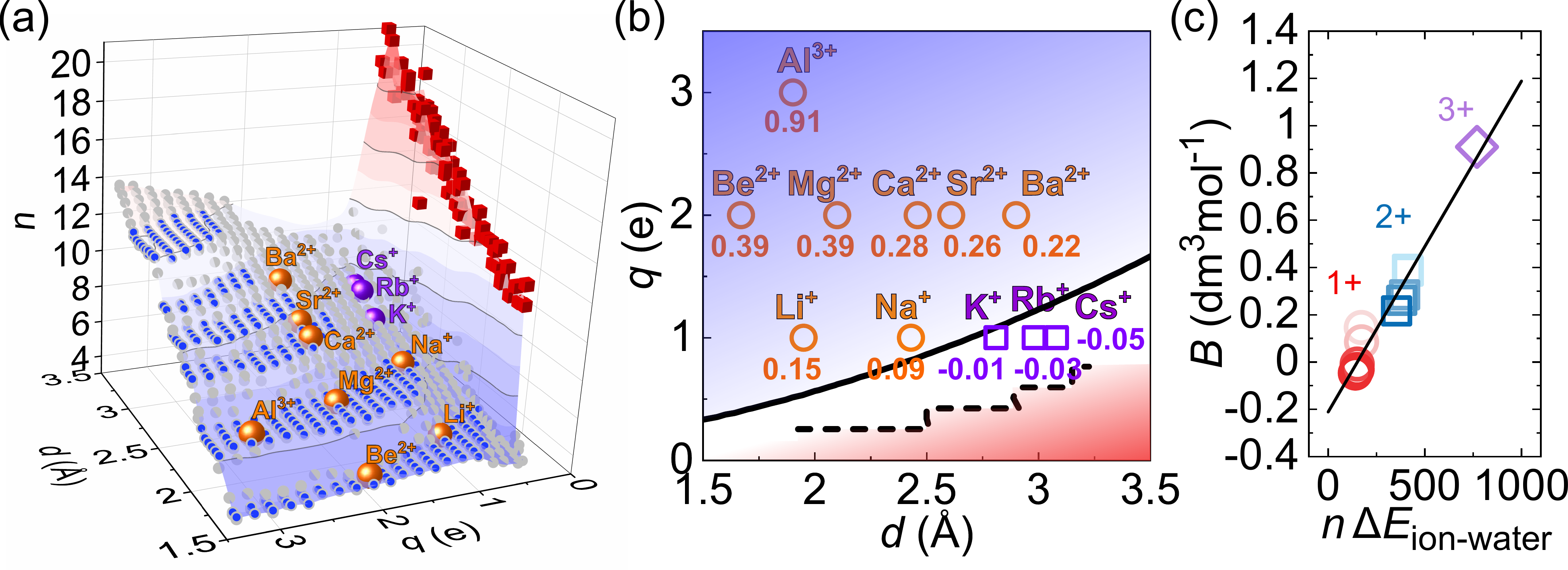}
	\end{center}
	\vspace{-7mm}
	\caption{The ``phase diagram'' of the hydration shell around spherical ions in water. (a) The coordination number $n$ of cationic ions obtained from experimental data (big spheres) and our model (small blue spheres and red cubes). (b) The $q-d$ ``phase diagram'' of aqueous ionic solutions. The experimental $B$-coefficient~\cite{jenkins1995} in the unit of dm$^3$mol$^{-1}$ at 298.15~K is shown for each ion. The solid line indicates the H-bonding energy scale, separating $B>0$ (orange circles) with $B<0$ (violet squares) ions. The dashed line represents the transition line from the Stokes (red shaded region) to the dielectric regime (blue shaded region). (c) The correlation between the $B$-coefficient experimentally estimated and the ion-water interaction energy in the hydration shell $n\Delta E_\mathrm{ion-water}$. Red circles and blue squares represent the data for alkali metal ions and alkaline metal ions, respectively, and diamond denotes the aluminium ion. Among the same symbol, the darker one represents the bigger ion.}
	\vspace{-0mm}
	\label{fig:phase}
\end{figure*}

To compare our simulation results with real solutions, we collected experimental data of the coordination number $n$ and ion-water distance $d$ (Table~\ref{table:data}) for alkali, alkaline, and aluminium ions. We can see from Fig.~\ref{fig:phase}(a)) that our model prediction agrees well with the experimental data. In Fig.~\ref{fig:phase}(b), we plot the experimental $B$-coefficient in the $q-d$ diagram. We can see in the diagram that the $\lambda_\mathrm{HB}$ line indeed separates kosmotropic ($B>0$) and chaotropic ($B<0$) ions. The structure maker/breaker classification has so far been made empirically based on the sign of the $B$-coefficients~\cite{mancinelli2007hydration,gallo2011ion}. The critical length $\lambda_\mathrm{HB}$ is determined by the competition of ion-water and water-water interactions, providing new microscopic insight into the sign of the $B$-coefficient. Furthermore, we find that the quantity $n\Delta E_\mathrm{ion-water}$ that measures the average ion-water interaction energy of $n$ hydrated water molecules correlates well with the experimental $B$-coefficients of ions (Figure~\ref{fig:phase}(c)). Our results confirm that both structure maker and breaker perturb the water structure, in agreement with previous scattering experiments~\cite{mancinelli2007hydration}.

\subsection*{Conclusion and outlook}
We have revealed a structural transition of the hydration shell in the $q-d$ space, accompanied by the crossover from the Stokes to the dielectric friction regime with an increase in the electric field of ions, which is realized for ions of stronger charge and/or smaller size.  This structural transition is sharp and accompanied by the significant reduction of hydrated water molecules and the development of translational order in the radial direction. 
There is another transition in the dielectric friction regime due to the competition between ion-water (monopole-dipole) electrostatic interaction and water-water H-bonding. This transition is accompanied by the dynamic transition of the solvent from the accelerated ($B<0$) to decelerated ($B>0$) behavior. We also find that this transition is characterized by a new characteristic length $\lambda_\mathrm{HB} (q)$, which marks the distance from the ion beyond which the electrostatic interaction does not significantly perturb the H-bonding.

We also discover the orientational ordering of water dipoles around ions in the hydration shell for a combination of $q$ and $d$ in the higher $q$ or lower $d$ regime. This local orientational ordering is accompanied by a drastic increase in hydration shell stability and the residence time of water in the hydration shell~\cite{lee2017}. This ordering and stabilization of the hydration shell occurs for a composite number of the hydrated water molecules and does not for a prime number. We have revealed that the physics behind this unique ordering in a spherical shell is essentially the same as that of the Thomson problem of electrons around positive nucleus: repulsion-induced ordering in the spherical shell. 

We have summarized our main results in Table \ref{table:summary}. All these discoveries were made possible only by a microscopic understanding of the structure of the water around the ions, which is not possible with the continuum theory. We hope that this work will contribute to the development of microscopic theories of solvation around spherical ions.

\begin{table*}[t!]
	\caption{Classification of the ion hydration states and dynamic characteristics.}
	\label{table:summary}
	\vspace{0mm}
	\begin{tabular}{|c||c|c||c|c|c|}
		\hline
		Electrical field  & Hydration  &Water structure & Friction & $B$ & Water dynamics  \\
		on ion surface& shell (HS) &around ion &  & coefficient & in HS \\
		\hline
		\hline
		weak  & thick & disordered  & viscous & $B \sim 0$ & comparable to bulk \\
		& & H-bonding in HS & (Stokes) & &  \\
		\cline{1-6}
		intermediate  & & low translational order &  &  & weakly accelerated \\
		$d>\lambda_{\rm HB}$ &  & weak radial alignment of dipoles &  & $B<0$ & (compared with bulk) \\
		&  & partial H-bonding in HS &  &  & \\
		\clineB{1-1}{3.5}\clineB{3-3}{3.5}\clineB{5-6}{3.5}
		&   & high translational order  & & &    \\
		& & strong radial alignment of dipoles & & & decelerated \\
		strong & & no-H-bonding in HS & viscous & & (compared with bulk)\\
		$d<\lambda_{\rm HB}$ & thin &- - - - - - - - - - - - - - - - - - - - - - -& + & $B>0$&   - - - - - - - - - - - - - - - - - - - -\\
		&  &  composite $n$: & dielectric & & longer residence time    \\
		& &  water polyhydral ordering &  && slower water diffusion    \\
		& &   &  & & (compared with prime $n$)    \\
		&    & - - - - - - - - - - - - - - - - - - - - - - -& && - - - - - - - - - - - - - - - - - - - -  \\
		&   & prime $n$:     &&& shorter residence time  \\
		& &    no polyhedral order &  &  & faster water diffusion  \\
		& &   &  & & (compared with composite $n$)    \\
		\hline
	\end{tabular}
\end{table*}

This work provides a simple physical picture of salt effects on water, based on the classical description of the electrostatic and VDW interactions. Our approach may be considered as a zero-order approximation for such complex effects. Higher-order effects, such as polarizability, many-body interactions, and nuclear quantum effects, would be necessary for a more precise and comprehensive description of ion-specific effects in aqueous solutions. Moreover, this study has considered only hydration around spherical ions, but solvation around molecular ions with complex shapes is an important topic for future research. We expect the characteristic length $\lambda_{\rm HB}$ and prime-number effect introduced here to be helpful in these cases as well.

This study was partly supported by Specially Promoted Research (JP20H05619 and JP25000002) and Scientific Research (A) (JP18H03675) from the Japan Society for the Promotion of Science (JSPS) and the Mitsubishi Foundation.

\clearpage

\bibliography{ion}

\clearpage

\setcounter{figure}{0}
\renewcommand{\figurename}{\textbf{Fig.}}
\renewcommand{\thefigure}{S\arabic{figure}}
\setcounter{table}{0}
\renewcommand{\tablename}{\textbf{Table}}
\renewcommand{\thetable}{S\arabic{table}}
\setcounter{equation}{2}

{\bf SUPPLEMENTARY MATERIAL}

\section*{Methods}
\label{sec:methods}

\vspace{-4mm}
\subsection*{Potentials for aqueous ionic solutions}
\vspace{-4mm}

It has been demonstrated that a realistic TIP4P/2005 water model~\cite{abascal2005} accurately describes the structural and dynamic properties of liquid water~\cite{vega2009}. Focusing on the water structure and dynamics, we employed a non-polarizable force field~\cite{zeron2019} that was newly developed for aqueous salt solutions, based on the TIP4P/2005 water model. This force field, compatible with the TIP4P/2005 water, shows reasonably good performance for modelling a series of aqueous salt solutions~\cite{zeron2019}.

The size and charge of the ions together determine the ion-solvent interactions. The intermolecular interaction $V$ between atoms $i$ and $j$ is given by the sum of VDW and Coulombic potentials:
\begin{equation}
	V\left(r_{ij}\right) = 4\epsilon_{ij}\left[\left(\frac{\sigma_{ij}}{r_{ij}}\right)^{12} - \left(\frac{\sigma_{ij}}{r_{ij}}\right)^{6}\right] + \frac{1}{4\pi\epsilon_0}\frac{q_i q_j}{r_{ij}},
	\label{eq:potential}
\end{equation}
where $r_{ij}$ is the distance between atoms $i$ and $j$, $\epsilon_{ij}$ is the energy scale of the VDW potential, $\sigma_{ij}$ is the VDW diameter, $\epsilon_0$ is the vacuum permittivity, and $q_i$ and $q_j$ are the charges of atoms $i$ and $j$.
The dielectric friction theory (equation~(\ref{eq:df})), which treats the solvent as the continuum media, has shown that the size and charge of an ion control dynamics of a given solvent. In order to investigate the dependence of salt effects on the ion size and charge, we employed the above-mentioned force field parameters for aqueous NaCl solution as the reference and then continuously tuned the VDW size and the charge of the cation, while the force field parameters being fixed for anion (Cl$^-$) and TIP4P/2005 water. Practically, we linearly modified the VDW parameter $\sigma$ between cation and water oxygen (chloride) as $\sigma_\mathrm{cation-O(Cl)} = \sigma_\mathrm{Na-O(Cl)} + k \cdot \delta \sigma$, where $k$ is an integer, $\delta \sigma$ is chosen to be 0.04~\AA~and $\sigma_\mathrm{Na-O(Cl)}$ is the VDW parameter in the original force field~\cite{zeron2019}.

In the original force field, the chloride ion has a reduced charge of -0.75 $e$ to include the polarizable effect effectively~\cite{kann2014}. Therefore, the cation charge is given by $q_\mathrm{cation} = k \cdot \delta q$, where $k$ is an integer and $\delta q$ is chosen as $\frac{0.85}{5} = 0.17~e$. Then, the number of ions is adjusted to ensure the charge neutrality of the system. In this way, we prepared potentials for 1332 model systems with cationic charge ranging from 0.17 to 3.4~$e$ that covers most of the alkali metal ions, the alkaline metal ions, and the IIIA-group metal ions.

\vspace{-4mm}
\subsection*{Potentials for aqueous ionic solutions with small fractional charges}
\vspace{-4mm}

We scanned 20 points in the charge space using the potential mentioned above with a resolution of $\delta q=0.17~e$ for a given cation size. In order to increase the resolution in the small-fractional-charge region, we prepared 25 new systems with the same VDW parameter $\sigma_\mathrm{cation-O} = 2.73$~\AA \,\ which is close to the VDW parameter for the sodium ion. In the new systems, the cation charge is varied from 0 to 1.2~$e$ with a resolution of $\delta q = 0.05~e$, and the anion has the same charge as the cation. The systems have the same number of cations and anions to keep the charge neutrality.

\vspace{-4mm}
\subsection*{Method for equilibrium simulations}
\vspace{-4mm}
We have performed high-throughput molecular dynamics simulations of 1332 model solutions containing cations with different sizes and charges. Simulations were performed using the Gromacs (v.4.6.7) simulation package~\cite{hess2008gromacs} with a time step of 2~fs. The system consists of 3456 water molecules, 10 cations, and some anions (from 2 to 40, determined by the charge neutrality condition) in a periodic cubic box. This corresponds to a dilute concentration of $\sim0.16$ mol/kg of the salt. We used the isothermal-isobaric $NPT$ ensemble for all the simulations. We employed the Nose-Hoover thermostat with a coupling time of 1.0~ps and an isotropic Parrinello-Rahman barostat with a coupling time of 2.0~ps to keep the temperature at 300~K and pressure at 1~bar. The Particle-Mesh Ewald method was applied for long-range electrostatic interactions. The VDW interactions and the Coulomb potential in real space were cut at 10~\AA. All the 1332 systems were first equilibrated for 0.3~ns and followed by a production run for 1~ns.

\vspace{-4mm}
\subsection*{Method for non-equilibrium simulations}
\vspace{-4mm}
A high-throughput scanning in the charge-size space revealed a structural transition of the ionic hydration shell in the small-fractional-charge region. In order to understand the mechanism of the structural transition, we performed non-equilibrium molecular dynamics for a system containing 3456 water molecules, 10 cations with $\sigma_\mathrm{cation-O} = 2.76$~\AA, and 8 chloride anion. For this system, the transition occurs at $q_s=0.4~e$. We first set the cation charge as $q=0.17~e$ below the transition threshold and equilibrated the system at 300~K and 1~bar for 1.0~ns. Then, 50 independent configurations were evenly sampled from 1-ns trajectories. We carried out 50 production runs using these configurations as the initial ones, at 300~K and 1~bar for 3~ps. 
Then, we jumped the charge of cations to $q=0.68~e$ at time 0 instantaneous to induce the transition. The configurations were sampled every 4~fs to follow the transition kinetics. The charge of anions was adjusted to ensure charge neutrality. The other simulation details are the same as the equilibrium simulations described above. Results obtained from these non-equilibrium simulations were averaged over the 50 trajectories to reduce statistical fluctuations.

\vspace{-4mm}
\subsection*{Structural analysis tools}
\vspace{-4mm}

To characterize the hydration structure around ions, we introduce the structural parameter $t$~\cite{errington2001} as $t=\frac{1}{\xi_\mathrm{c}} \int_{0}^{\xi_\mathrm{c}} \vert g\left(\xi\right)-1\vert \mathrm{d} \xi$, where $g$ is the cation-oxygen RDF, $\rho$ is the number density of water, $\xi_\mathrm{c}=2.843$ is a cut-off distance. The parameter $t$ measures the degree of translational order of solvent molecules around a cation in the radial direction.

\vspace{-4mm}
\subsection*{Dynamic analysis tools}
\vspace{-4mm}

The diffusion coefficient $D_1$ characterizes the speed of diffusive motion of a hydrated water molecule in the hydration shell. The mean-squared displacement of water molecule $i$ in the hydration shell of ion $j$ at $t=0$ can be calculated from the trajectory as $\mathrm{MSD} \left(t\right) = \langle \left[\vec{r}_i\left(t\right)-\vec{r}_i\left(0\right)\right]^2 \delta \left[\vert \vec{r}_i\left(0\right) - \vec{r}_j\left(0\right) \vert - r_1 \right] \rangle$, where $r_1$ is the characteristic size of the hydration shell determined as the first minimum position of the cation-oxygen RDF $g(r)$, $\delta\left(x\right)=1$ if $x\leqslant0$ and $\delta\left(x\right)=0$ otherwise, $\langle  \cdots \rangle$ denotes the ensemble average. The diffusion coefficient $D_1$ is then obtained, using the Einstein relation, $\mathrm{MSD} = 6D_1t$, where an upper bound $\mathrm{MSD}\leqslant 16$~\AA\,\ is set to ensure the locality of $D_1$. 

The residence time $\tau_\mathrm{R}$ of a water molecule in the hydration shell can be characterized by the continuous time correlation function, $P\left(t\right) = \langle \frac{p\left(t\right)p\left(0\right)}{p\left(0\right)} \rangle$, where $p\left(t\right)=1$ if a water molecule that is in the hydration shell of an ion at $t=0$ continuously stays in the hydration shell of the same ion at time $t$; otherwise, $p\left(t\right)=0$. Then, the residence time can be extracted from $P\left(t\right)$, using the following equation, $P\left(\tau_\mathrm{R}\right) = e^{-1}$. Here, we regard a water molecule to remain in the hydration shell if it is within a distance $r_1$ from the central ion.

\section*{Supplementary Figures and Tables}

\begin{figure*}[h!]
	\begin{center}
		\includegraphics[width=16.cm]{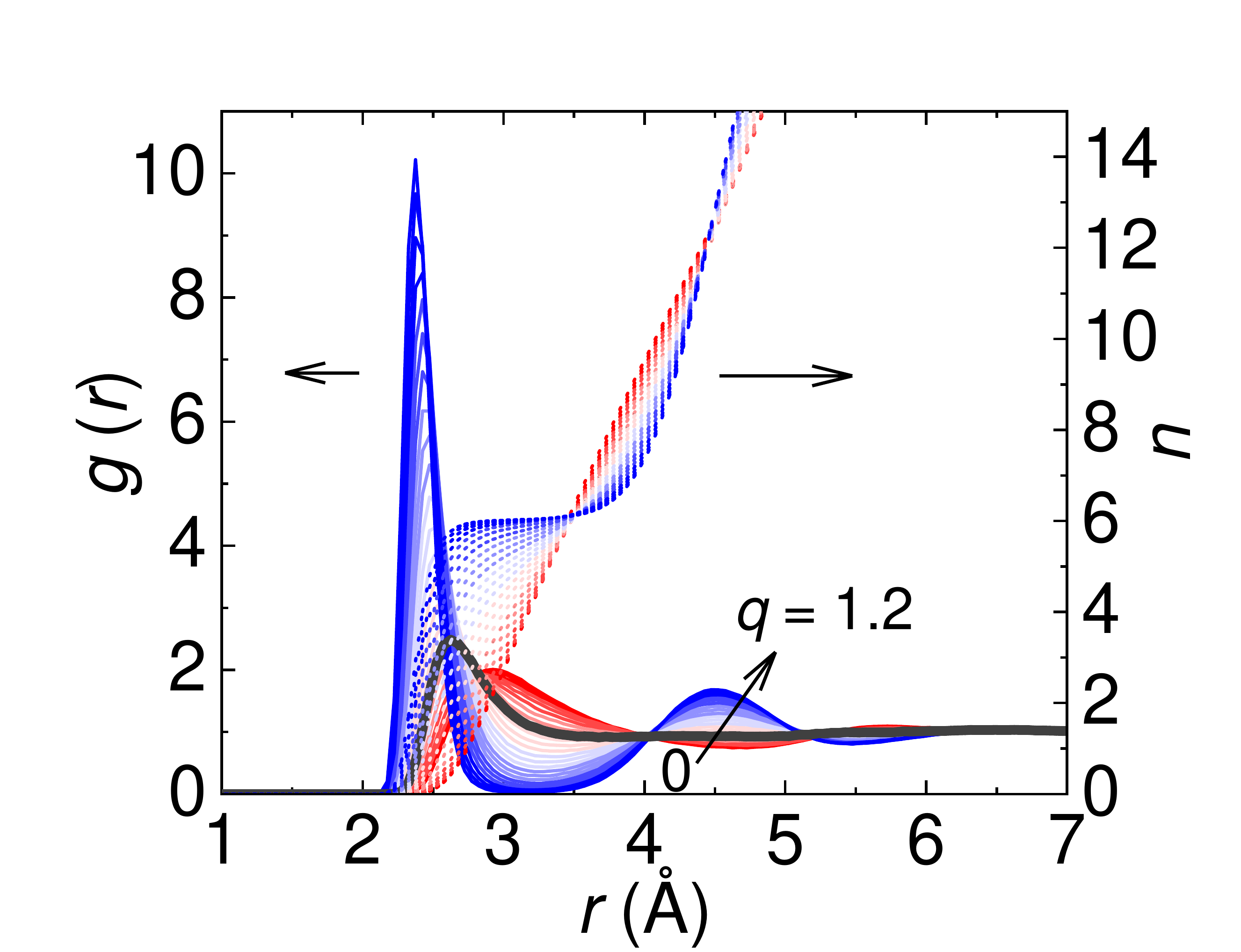}
	\end{center}
	\vspace{-0mm}
	\caption{Ion-water (cation-oxygen) radial distribution function (RDF) in aqueous ionic solutions along the green dashed line in Fig.~\ref{fig:transition}(a). In the solutions, the cation charge is varied from $q=0$ to $q=1.2~e$ with $\delta q=0.05~e$, while the VDW size of cations is fixed to $\sigma_\mathrm{cation-O} = 2.73$~\AA. The RDF at the structural transition ($q_s=0.4~e$) is highlighted by the black thick line. The coordination number $n$ is shown by dotted curves (see the right axis for the value).}
	\vspace{-0mm}
	\label{fig:rdfq}
\end{figure*}

\begin{figure*}[h!]
	\begin{center}
		\includegraphics[width=18.6cm]{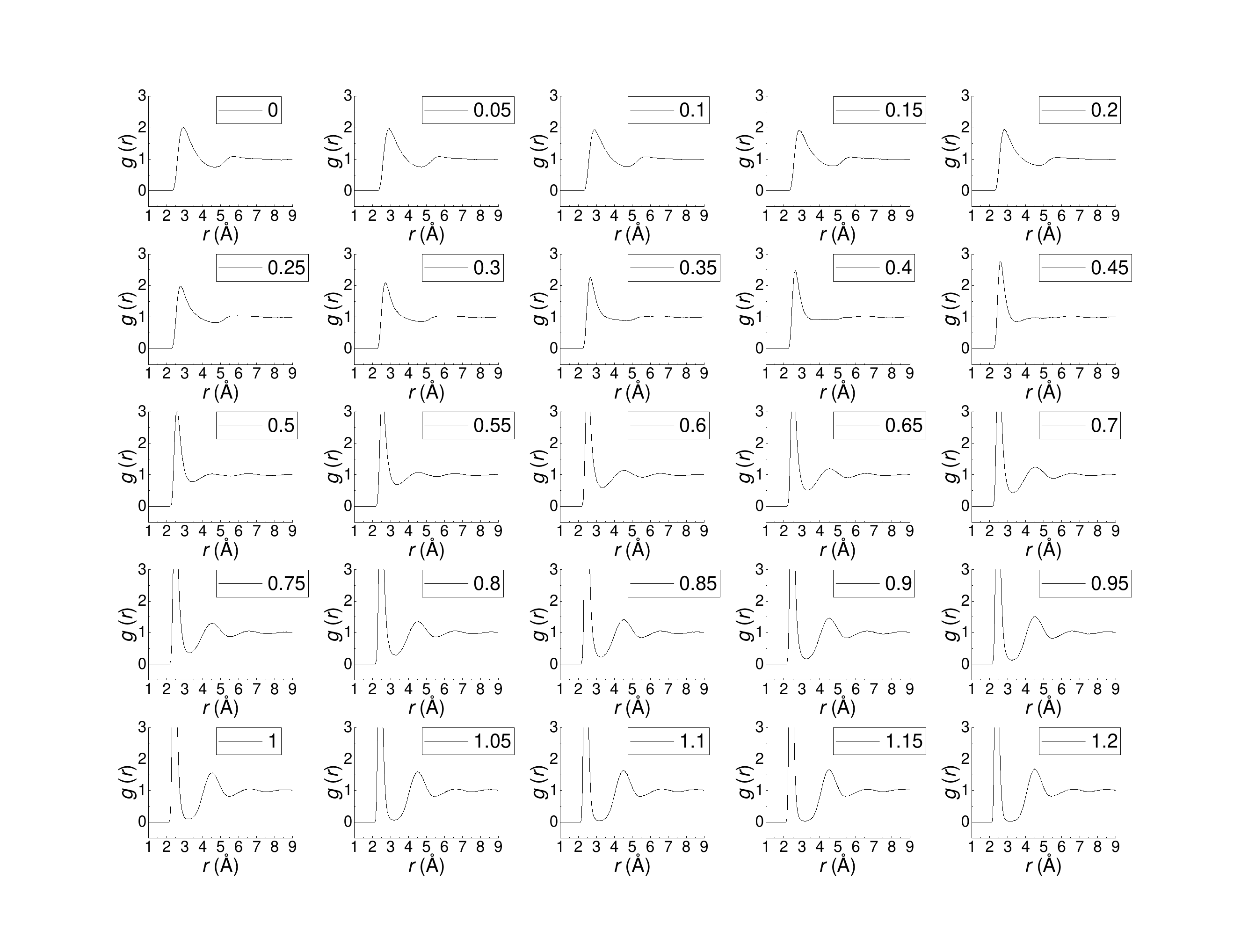}
	\end{center}
	\vspace{-7mm}
	\caption{Individual ion-water (cation-oxygen) RDF in aqueous ionic solutions along the green dashed line in Fig.~\ref{fig:transition}(a). The results are the same as Fig.~\ref{fig:rdfq}, but displayed separately.}
	\vspace{-0mm}
	\label{fig:rdfq1}
\end{figure*}

\begin{figure*}[h!]
	\begin{center}
		\includegraphics[width=16cm]{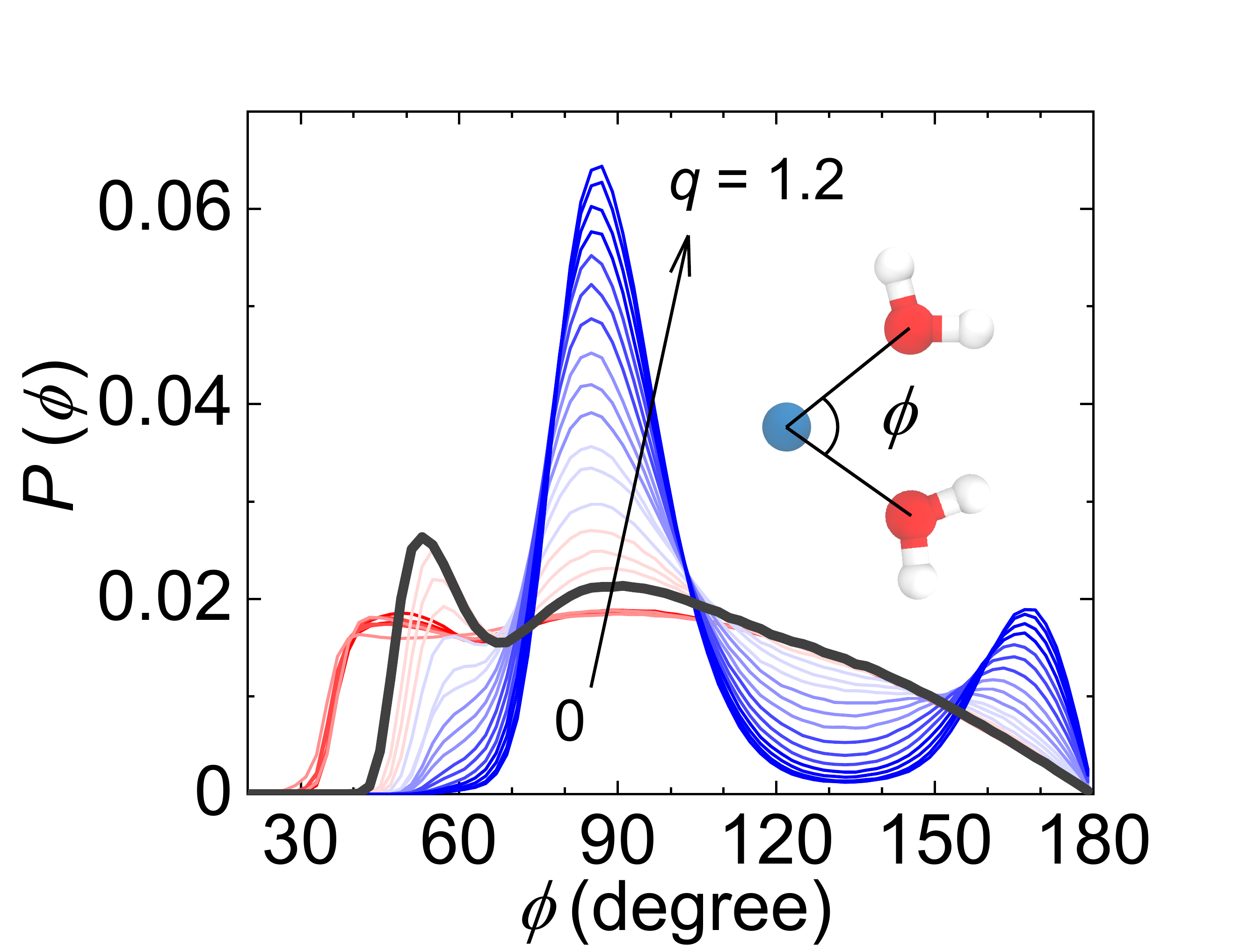}
	\end{center}
	\vspace{-0mm}
	\caption{Probability distribution function of the oxygen-cation-oxygen angle $\phi$, $P(\phi)$, in aqueous ionic solutions along the green dashed line in Fig.~\ref{fig:transition}(a). The angle $\phi$ is defined as the angle formed between the cation and two neighboring water oxygen atoms in the hydration shell (see inset). We varied the cation charge from $q=0$ to $q=1.2~e$ with a step of $\delta q=0.05~e$, whereas the VDW size of cations was fixed to $\sigma_\mathrm{cation-O} = 2.73$~\AA. $P(\phi)$ at the structural transition (i.e., at $q_s=0.4~e$) is highlighted by the black thick line.}
	\vspace{-0mm}
	\label{fig:angleq}
\end{figure*}

\begin{figure*}[h!]
	\begin{center}
		\includegraphics[width=16.cm]{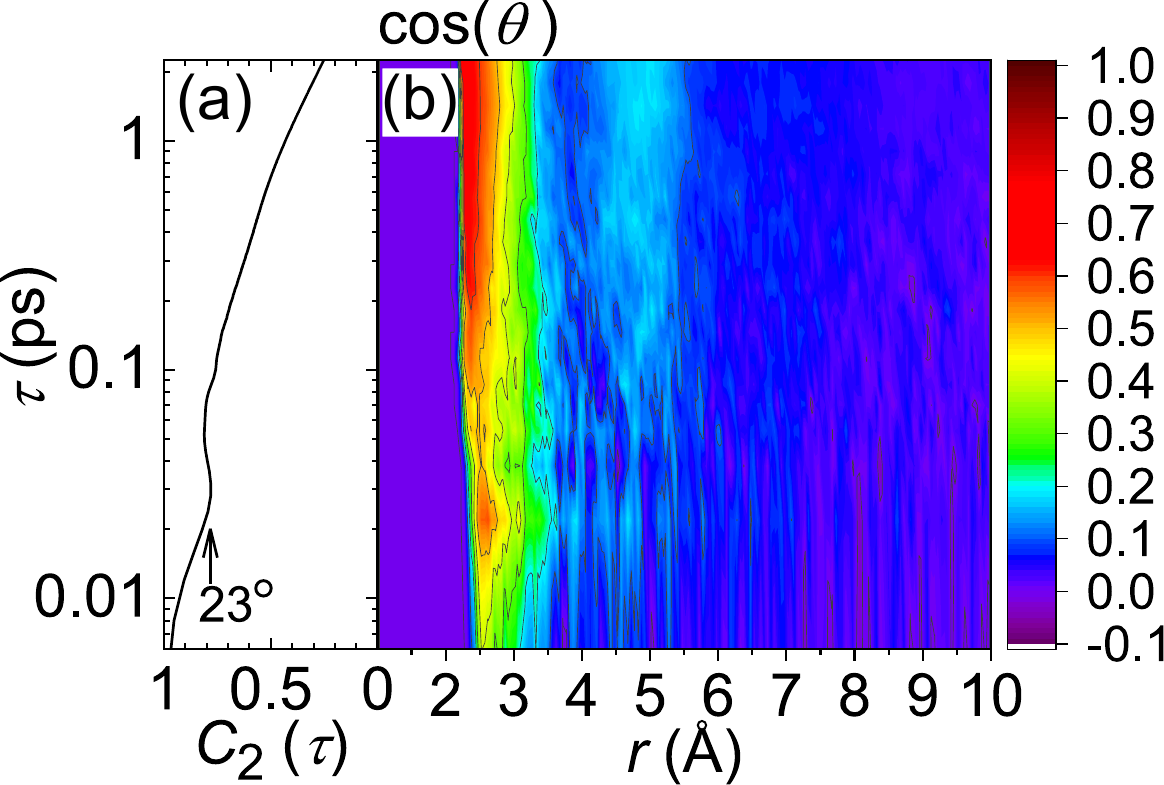}
	\end{center}
	\vspace{-0mm}
	\caption{Non-equilibrium structural transformation process in response to the instantaneous jump of cationic charge at time $\tau=0$. (a) Schematic illustration of 
	the definition of angle $\theta$, which is the angle formed by the water dipole and the ion-oxygen vector. (b) The temporal decay of the rotational time correlation function $\left(C_2(\tau)=1.5\langle\vec{\mu}(\tau)\vec{\mu}(0)\rangle^2-0.5\right)$ of the water dipole moment $\vec{\mu}$ as a function of time $\tau$ in bulk TIP4P/2005 water. (c) Time evolution of $\langle \cos \theta \rangle$ for water molecules in the hydration shell of cations in response to the instantaneous jump of the cation charge from $q=0.17$ to $0.68~e$ at $\tau=0$. The VDW size of cations was fixed to $\sigma_\mathrm{cation-O} = 2.76$~\AA. The color bar represents the value of $\langle \cos \theta \rangle$.}
	\vspace{-0mm}
	\label{fig:theta}
\end{figure*}

\begin{figure*}[h!]
	\begin{center}
		\includegraphics[width=18cm]{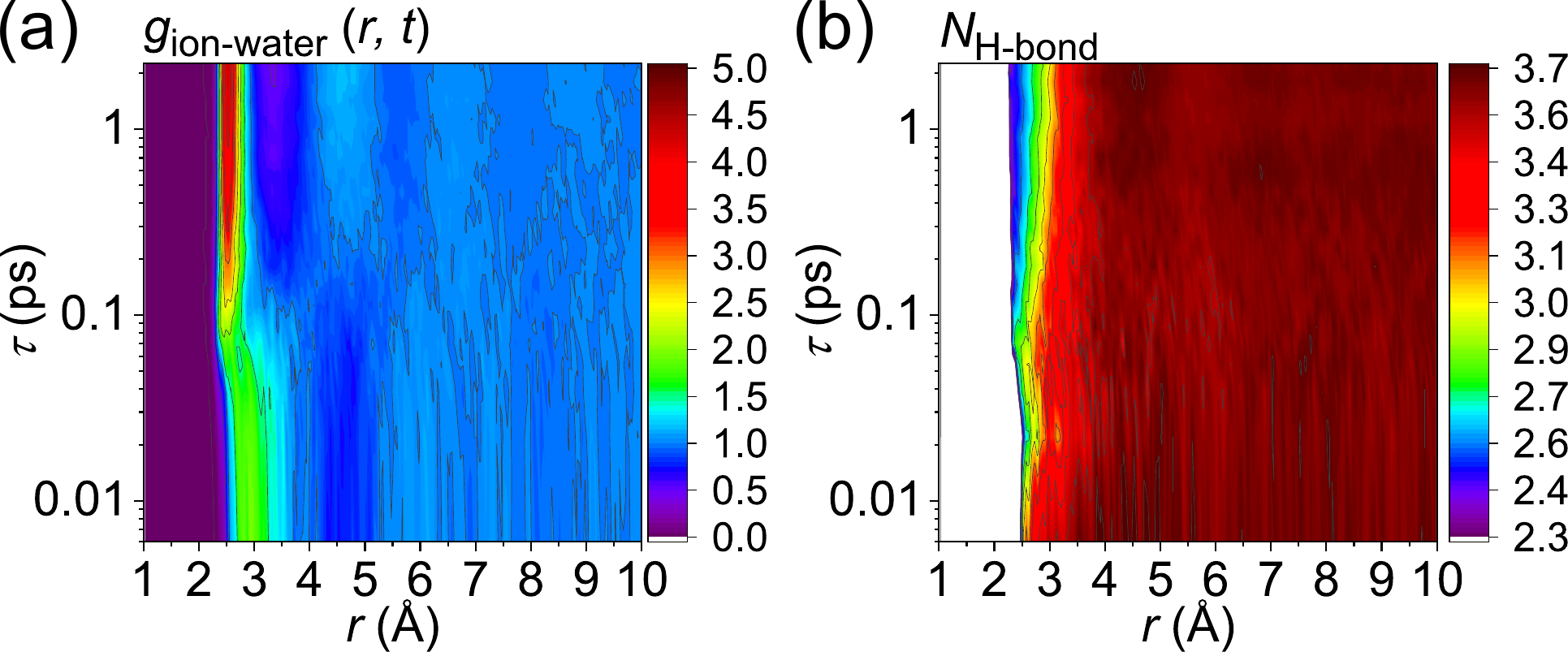}
	\end{center}
	\vspace{-0mm}
	\caption{Non-equilibrium structural transformation process in response to the instantaneous jump of cationic charge at time $\tau=0$. Time evolution of the ion-water RDF $g_\mathrm{ion-water} (r, \tau)$ (a) and the H-bond number $N_\mathrm{H-bond}$ per water molecule (b) in response to the instantaneous jum of cationic charge from $q=0.17$ to $0.68~e$ at $\tau=0$. The color bars in panels a and b represent the height of $g_\mathrm{ion-water}$ and 
the value of $N_\mathrm{H-bond}$, respectively.	}
	\vspace{-0mm}
	\label{fig:rdfnhb}
\end{figure*}

\begin{figure*}[h!]
	\begin{center}
		\includegraphics[width=16.cm]{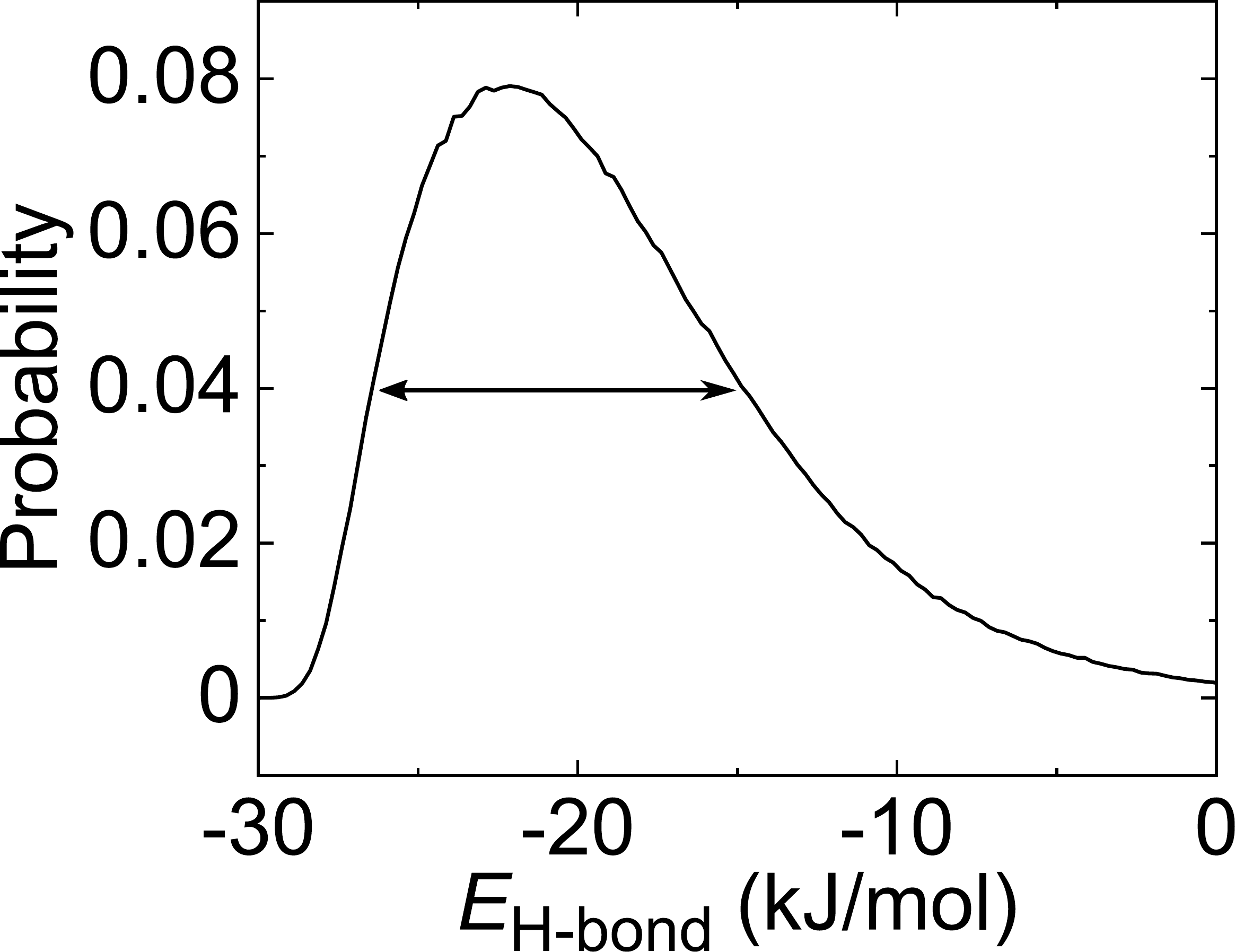}
	\end{center}
	\vspace{-0mm}
	\caption{Inter water H-bonding energy in bulk water. Probability distribution function of the H-bonding energy $E_\mathrm{H-bond}$ in bulk TIP4P/2005 water at 300 K and 1 bar. The full width at half maximum, as shown by the arrow, indicates the range of fluctations of the H-bonding energy in bulk water, which corresponds to the width of the yellow band in Fig.~\ref{fig:hbond}(b).}
	\vspace{-0mm}
	\label{fig:hbenergy}
\end{figure*}

\begin{figure*}[h!]
	\begin{center}
		\includegraphics[width=18.cm]{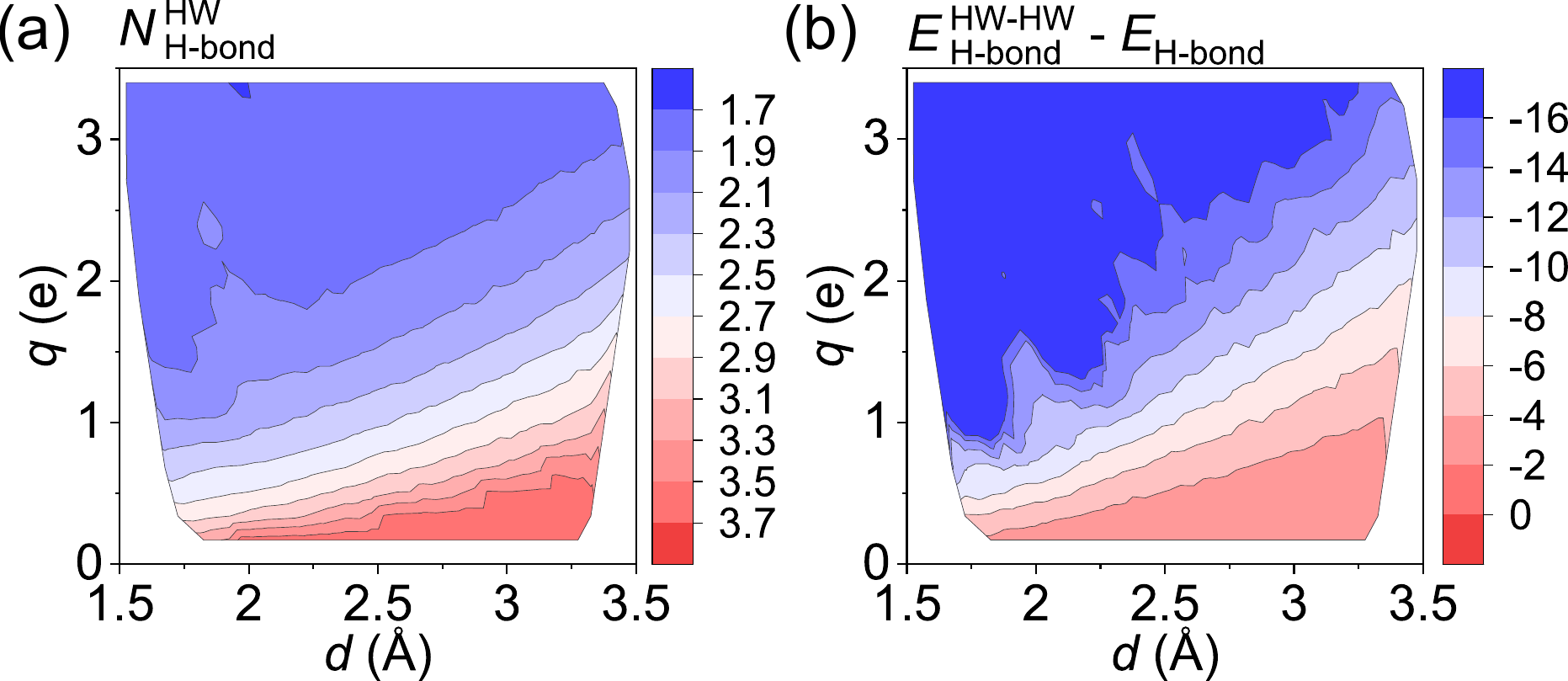}
	\end{center}
	\vspace{-0mm}
	\caption{Inter water H-bonding in hydration shell. (a) Average number of H-bonds formed for a water molecule in the hydration shell of cation, $N_{\rm H-bond}^{\rm HW}$. Here we consider all H-bonds formed for a hydrated water with other waters within and outside the same hydration shell. This is different from $N_\mathrm{H-bond}^\mathrm{HW-HW}$, which only counts the number of H-bonds between a hydrated water molecule and other water molecules within the same hydration shell (see Fig.~\ref{fig:hbond}(c)). 
	The color bar represents the value of $N_{\rm H-bond}^{\rm HW}$. Note that in bulk TIP4P/2005 water, each water molecule forms 3.7 H-bonds on average at 300~K and 1~bar. (b) The energy difference between the H-bond formed by two hydrated water in the hydration shell of the same cation and the H-bond formed in bulk water, $E_{\rm H-bond}^{\rm HW-HW}-E_{\rm H-bond}$. The color bar represents the value of $E_{\rm H-bond}^{\rm HW-HW}-E_{\rm H-bond}$ in the unit of kJ/mol.}
	\vspace{-0mm}
	\label{fig:nhb}
\end{figure*}

\begin{figure*}[h!]
	\begin{center}
		\includegraphics[width=12.cm]{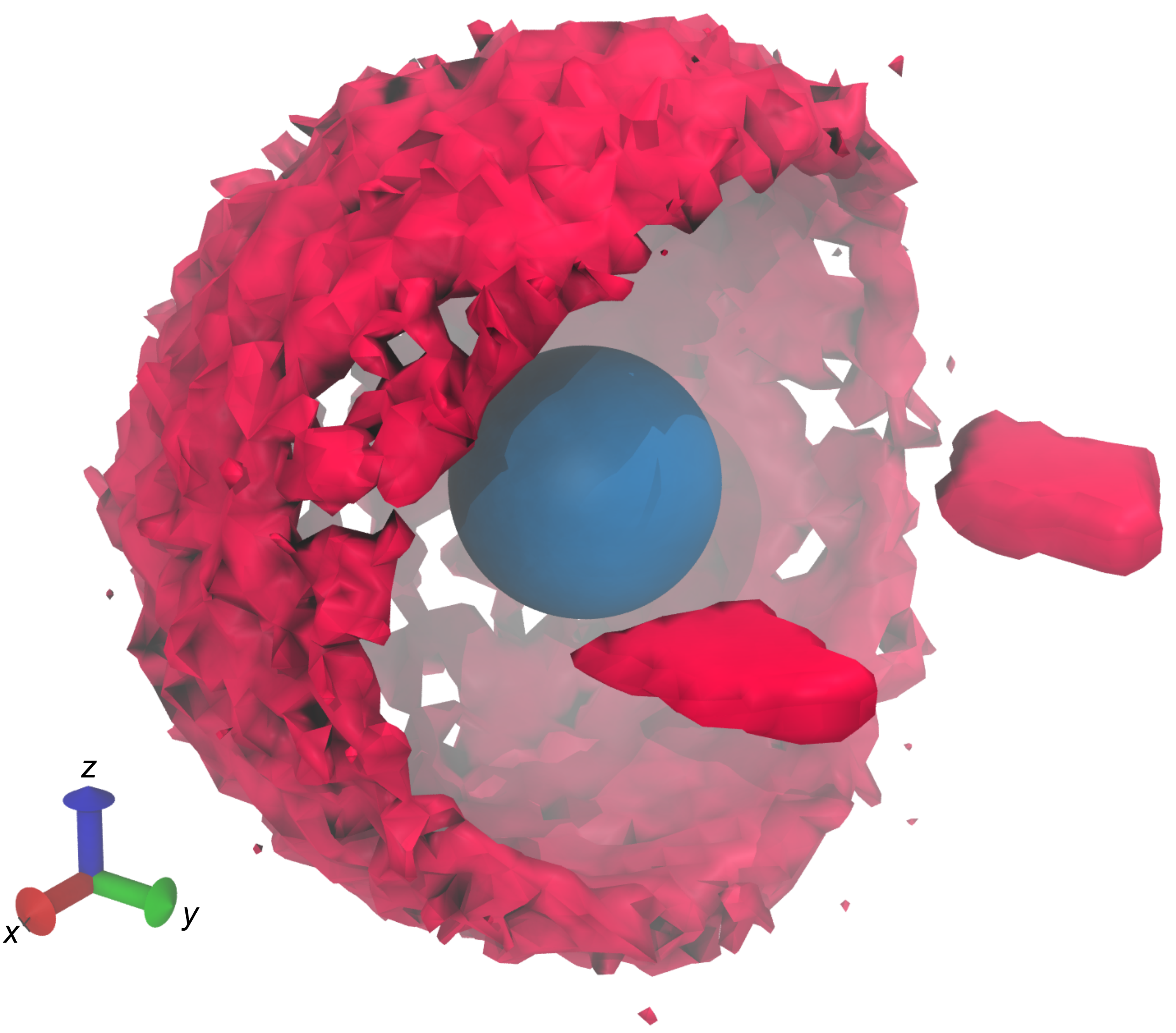}
	\end{center}
	\vspace{-0mm}
	\caption{Orientational symmetry of a typical unstable hydration shell. Spatial distribution of water molecules around a cation with the coordination number of $n=6$. We can see that the distribution has a homogeneous angular distribution without the breakdown of orientational symmetry. 
We note that this particular state point (the cation charge $q=0.51~e$ and the size of the hydration shell $d=2.6$~\AA) is located outside of the plateau region of Fig.~\ref{fig:symmetry}(a), and the hydration shell is not stable. Two neighboring water molecules are selected to determine the $xyz$ axes (two separated clouds on the right-handed side). The size of the ion in the centre does not represent the actual size of the ion.}
	\vspace{-0mm}
	\label{fig:sdf}
\end{figure*}

\clearpage

\begin{table*}[h!]
	\caption{The radius $d$ of hydration shell and the coordination number $n$ of cations in aqueous solutions obtained from experimental measurements.}
	\label{table:data}
	\vspace{0mm}
	\begin{tabular}{|c|c|c|c|c|c|}
		\hline 
		Cation & Salt & $d$~(\AA) & $n$ & Method & reference \\ 
		\hline
		Li$^+$ & LiCl & 1.95 & 4.0 & X-ray and Neutron & Ref.~\cite{narten1973} \\ 
		\hline
		Na$^+$ & NaI & 2.43 & 6.0 & X-ray & Ref.~\cite{mahler2012} \\ 
		\hline
		K$^+$ & KI & 2.81 & 7.0 & X-ray & Ref.~\cite{mahler2012} \\ 
		\hline
		Rb$^+$ & RbI & 2.98 & 8.0 & X-ray & Ref.~\cite{mahler2012} \\ 
		\hline
		Cs$^+$ & CsI & 3.081 & 8.0 & X-ray & Ref.~\cite{mahler2012} \\ 
		\hline
		Be$^{2+}$ & BeCl$_2$ & 1.67 & 4.0 & X-ray & Ref.~\cite{yamaguchi1986} \\ 
		\hline
		Mg$^{2+}$ & MgCl$_2$ & 2.1 & 6.0 & X-ray & Ref.~\cite{caminiti1979} \\ 
		\hline
		Ca$^{2+}$ & CaCl$_2$ & 2.46 & 8.0 & X-ray & Ref.~\cite{jalilehvand2001} \\ 
		\hline
		Sr$^{2+}$ & SrCl$_2$ & 2.61 & 8.3 & X-ray & Ref.~\cite{parkman1998} \\ 
		\hline
		Ba$^{2+}$ & BaCl$_2$ & 2.9 & 9.5 & X-ray & Ref.~\cite{albright1972} \\ 
		\hline
		Al$^{3+}$ & AlCl$_3$ & 1.902 & 6.0 & X-ray & Ref.~\cite{caminiti1979order} \\ 
		\hline	
	\end{tabular} 
\end{table*}

\end{document}